\documentclass[conference]{IEEEtran}

\usepackage{cite}
\usepackage{multirow}
\usepackage{booktabs}
\usepackage{psfrag}
\usepackage[utf8]{inputenc}
\usepackage{color}
\usepackage[dvips]{graphicx}
\usepackage{amsmath}
\usepackage{amssymb}
\usepackage{algorithm}
\usepackage{algpseudocode}
\usepackage{array}
\usepackage{eurosym}
\begin{document}

\title{Distributed AC-DC Optimal Power Flow in the European Transmission Grid with ADMM}

\author{%
	\IEEEauthorblockN{\underline{Nico Huebner}, Yannick Rink, Michael Suriyah, Thomas Leibfried\\}
	\IEEEauthorblockA{Institute of Electrical Energy Systems and High Voltage Technology \\ Karlsruhe Institute of Technology (KIT), Germany\\
		\{nico.meyer-huebner, yannick.rink, michael.suriyah, thomas.leibfried\}@kit.edu}
}



\maketitle

\begin{abstract}
Distributed or multi-area optimal power flow appears to be promising in order to cope with computational burdens in large-scale grids and without the regional system operators losing control over their respective areas. However, algorithms are usually tested either in small test cases or single countries. We present a realistic case study in the interconnected European transmission grid with over 5000 buses and 315\:GW load. The grid is partitioned into 24 areas which correspond to the respective single countries. We use a full alternating current model and integrate multi-terminal direct current systems. The decomposed problem is solved via a modified Alternating Direction of Multipliers Method (ADMM), in which the single countries only exchange border node information with each neighbor. In terms of generation costs, the solution of the distributed optimal power flow problem deviates only 0.02\% from a centrally computed one. Consensus between all regions is reached before 200 iterations, which is remarkable in such a large system.
\end{abstract}


\IEEEpeerreviewmaketitle

\section{Introduction}

In Europe, coupled market areas and the increasing penetration of renewable energy sources regularly lead to high cross-border power transfers and require frequent interactions between neighboring transmission system operators (TSO). For instance, congestion management and generator redispatch become more and more a matter of international rather than national interest. Furthermore, cost-efficiency should be aimed at on a European level. A cheap local measure should be avoided if it results in large additional costs for the neighboring TSO.

The most important economic optimization which respects electrical grid constraints is the Optimal Power Flow (OPF), which was introduced by Carpentier~\cite{Carpentier1962} and studied to large extents in the last decades~\cite{Momoh1999a,Momoh1999b}. However, in its original form, it requires information and control over the entire system. 
Motivated by the increasing complexity of the power system and the resulting computational burdens, the interest in distributing the optimization grew. An early overview on distributed OPF algorithms can be found in~\cite{Kim2000} and the most recent developments are examined in~\cite{Molzahn2017}. 

The most popular branches to tackle the non-convex alternating current (AC) OPF problem are the Optimality Condition Decomposition (OCD)~\cite{Conejo2002} and the Alternating Direction of Multipliers Method (ADMM)~\cite{Boyd2011}. In OCD technique, information among regions is exchanged after each iteration of an interior point method. Contrarily, in ADMM, each agent solves its sub-problem to optimality and then exchanges information. 

One main question is whether or not algorithms are applicable to real-world problems. Cases with up to 300 buses are solved in~\cite{Nogales2003,Hug2009,Erseghe2014,Erseghe2015,Gavriluta2016,Mhanna2017,Engelmann2018}. In~\cite{Liang2014,Guo2017,Shin2019}, the Polish transmission system with over 3000 buses is used. Additionally, \cite{Shin2019} shows a case with over 6000 buses of the French system. However, \cite{Liang2014} and~\cite{Shin2019} use network reduction methods to reduce the system size. In~\cite{Huang2002}, a network with over 8000 buses is solved, however, there is no information on grid type or load. All of the above works have in common at least one of the following drawbacks:
\begin{itemize}
	\item small system size
	\item single country and therefore comparatively small system load
	\item mathematical clustering of buses disregarding geopolitical or regulatory boundaries.
\end{itemize}

In this work, we overcome those drawbacks and present a realistic case study of the European transmission system with over 5000 buses and 315\:GW load in 24 countries. We use a modified ADMM, where penalties are imposed with different weights depending on the coupling variable as presented in~\cite{MeHue2019,Huebner2019,Huebner2019_Diss}. While sub-optimality gaps of 0.54\:\%~\cite{Liang2014}, 0.43\:\%~\cite{Guo2017} or 1\:\%~\cite{Shin2019} are achieved for large-scale systems in literature, the modified ADMM in this work reduces the gap to 0.02\:\%.
Furthermore, we extend the standard OPF by integrating high voltage direct current (HVDC) transmission systems in voltage source converter (VSC) technology. As shown in~\cite{MeHue2018}, converter power set point optimization will play an important role in cost minimization and~\cite{MeHue2019,Huebner2019_Diss} show that HVDC systems can efficiently be included in distributed optimization as well.
Thus, the main contributions of this paper are:
\begin{itemize}
	\item first large-scale distributed AC-DC OPF 
	\item first large-scale case study with geopolitical node clustering with by far the highest system load
	\item improvement of optimality gap by one order of magnitude with recently proposed modified ADMM.
\end{itemize}

The paper is organized as follows. In Chapter~\ref{ch:OPF}, we present the centralized formulation for an AC-DC OPF. In Chapter~\ref{ch:Distribute}, the decoupling of regions and the ADMM algorithm are presented. The network model and simulation results are shown in Chapter~\ref{ch:Results}, and we draw a conclusion in Chapter~\ref{ch:Conclusion}.

\section{AC-DC Optimal Power Flow}\label{ch:OPF}
Consider an electrical network with a total of $N$ nodes, which are collected in the set $\mathcal N = \{1,...,N\}$. It consists of alternating current (AC) nodes $\mathcal N^\text{AC}\subseteq\mathcal N$ and direct current (DC) nodes $\mathcal N^\text{DC}\subseteq\mathcal N$. Two AC nodes are connected via AC lines, two DC nodes are connected via DC lines, and an AC and a DC node are connected via an AC-DC converter.
The centralized formulation of the full AC-DC OPF problem is written as
\begin{subequations}
	\begin{align}
	\underset{\substack{|V_{\text{AC}}|, \angle V_{\text{AC}},  \\ P_{\text{G}}, Q_{\text{G}}, \\V_{\text{DC}},P_{\text{C}}, Q_{\text{C}}}}{\text{minimize}}  & \quad \sum_{l \in \mathcal N^\text{AC}} \bigl(b_{\text{G}} P_{\text{G,}l}+a_{\text{Q}} (Q^2_{\text{G,}l}+ Q^2_{\text{C,}l}) \bigr) \label{eq:objective}\\
	\text{subject to}	&\quad \operatorname{Re}\bigl(V_{\text{AC,}i}\sum_{l \in \mathcal N^\text{AC}}(Y_{\text{AC,}il}V_{\text{AC,}l})^*\bigr) = \notag \\
	& \hspace{90pt} P_{\text{G,}i}+P_{\text{C,}i}-P_{\text{D,}i} \label{eq:Pbalance} \\
	&\quad \operatorname{Im}\bigl(V_{\text{AC,}i}\sum_{l \in \mathcal N^\text{AC}}(Y_{\text{AC,}il}V_{\text{AC,}l})^*\bigr) = \notag \\
	& \hspace{90pt} Q_{\text{G,}i}+Q_{\text{C,}i}-Q_{\text{D,}i} \label{eq:Qbalance}\\
	&\quad \underline{V}_{\text{AC,}i} \leq |V_{\text{AC,}i}| \leq \overline{V}_{\text{AC,}i} \label{eq:ACvoltage}\\
	&\quad V_{\text{AC,}m}=1.0\text{\:pu\:}\angle 0^\circ  \label{eq:ACslack}\\
	&\quad \underline{P}_{\text{G,}i} \leq P_{\text{G,}i} \leq \overline{P}_{\text{G,}i} \label{eq:Pgen}\\ 
	&\quad \underline{Q}_{\text{G,}i} \leq Q_{\text{G,}i} \leq \overline{Q}_{\text{G,}i}  \label{eq:Qgen}\\
	&\quad V_{\text{DC,}j}\sum_{l \in \mathcal N^\text{DC}}(Y_{\text{DC,}jl}V_{\text{DC,}l}) = -(P_{\text{C,}j}+P_{\text{CL,}j}) \label{eq:DCbalance}\\
	&\quad \underline{V}_{\text{DC,}j} \leq V_{\text{DC,}j} \leq \overline{V}_{\text{DC,}j} \label{eq:DCvoltage}\\
	&\quad  V_{\text{DC,}n}=1.0\text{\:pu} \label{eq:DCslack}\\
	&\quad P_{\text{C,}i}+Q_{\text{C,}i} \leq \overline{S}_{\text{C,}i}^2  \label{eq:VSClimit}\\
		&\quad \forall i\in\mathcal{N}^\text{AC},\forall j\in\mathcal{N}^\text{DC},\forall m\in\mathcal{N}^\text{AC}_\text{ref},\forall n\in\mathcal{N}^\text{DC}_\text{ref}\notag.
	\end{align}\label{eq:OPF}
\end{subequations}

The optimization variables are the following: $(|V_{\text{AC}}|, \angle V_{\text{AC}})$ are magnitude and angle of all AC node voltages; generator active and reactive power is denoted with $(P_{\text{G}},Q_{\text{G}})$. The DC system model requires DC node voltages $V_{\text{DC}}$ and active and reactive power output of an AC-DC converter are modeled with $(P_{\text{C}},Q_{\text{C}})$. Each variable can be assigned to a certain node, for example, $P_{\text{C,}i}$ is a converter connected to node $i$. For the sake of readability, we assume at most one unit per type at the same node.

We use objective (\ref{eq:objective}) which minimizes active losses if we choose the same $b_\text{G}$ for all generators. Furthermore, we allow for a small weight to reactive power injections in order to regularize the problem and thus improve its numerical condition. Technically, this is motivated by keeping reactive power injections small. The parameter $a_{\text{Q}}$ is the quadratic coefficient for all reactive injections. 

The power balance of active and reactive power for an AC node is given by~(\ref{eq:Pbalance}) and~(\ref{eq:Qbalance}), respectively, where $Y_{\text{AC,}il}$ is the $il$-th entry of the complex AC bus admittance matrix. Furthermore, ($P_{\text{D,}l}, Q_{\text{D,}l}$) are the active and reactive power demand at node~$i$. Voltage magnitudes are limited by~(\ref{eq:ACvoltage}) and both magnitude and angle are fixed for one node per synchronous area~(\ref{eq:ACslack}). Reference nodes are collected in $\mathcal{N}^\text{AC}_\text{ref}\subset\mathcal{N}^\text{AC}$ and $\mathcal{N}^\text{DC}_\text{ref}\subset\mathcal{N}^\text{DC}$, respectively.
Generator power outputs must satisfy operational upper and lower limits~(\ref{eq:Pgen})-(\ref{eq:Qgen}).
Equivalently to the AC side, the active power balance of a DC node is given by~(\ref{eq:DCbalance}) and $Y_{\text{DC,}jl}$ is the $jl$-th entry of the real DC bus admittance matrix.
Again, voltages are limited by~(\ref{eq:DCvoltage}) and the voltage magnitude of one node is fixed to 1\:pu (\ref{eq:DCslack}).

The converter model is a simplified VSC model, where the active power transfer between AC and DC side is coupled via $P_{\text{C}}$, which is part of both AC power balance~(\ref{eq:Pbalance}) and DC power balance~(\ref{eq:DCbalance}). Here, a positive value denotes a power flow from DC to AC. The withdrawn power from the DC side is augmented by a loss term $P_{\text{CL}}$, which is quadratically dependent on the apparent power $S_{\text{C}}$. 
Lastly, we limit the apparent power injection on the AC side with~(\ref{eq:VSClimit}).

We may reformulate problem (\ref{eq:OPF}) to
\begin{subequations}
	\begin{align}
	\underset{x}{\text{minimize}}  & \quad F(x) \\
	\text{subject to}	&\quad h(x)\leq 0.
	\end{align}\label{eq:OPF_2}
\end{subequations}
Note that for later purposes and readability reasons, equality constraints have been integrated into the inequality constraints set $h(x)\leq 0$.

\section{Distributed Optimization}\label{ch:Distribute}
We form a regional formulation for the AC-DC OPF. We define $R$ regions in $\mathcal{R}=\{1,...,R\}$. Each region may contain both AC and DC nodes. Thus, a boundary between two regions runs between two nodes, crossing a tie line. The decoupling of such a tie line is described in the following section. Then, problem~(\ref{eq:OPF_2}) is brought into a regional form and lastly, the implemented algorithm is described shortly.

\subsection{Decoupling of Inter-Regional Connectors}

\begin{figure}
	\centering
	\psfrag{F}[l][l]{\small$\begin{aligned}&P_{\text{G,}n}\\(&Q_{\text{G,}n})\end{aligned}$}
	\psfrag{A}[r][r]{\small$\begin{aligned}&P_{\text{G,}m}\\(&Q_{\text{G,}m})\end{aligned}$}
	\psfrag{G}[c][c]{\small$m$}
	\psfrag{H}[c][c]{\small$n$}
	\psfrag{I}[c][c]{\small$i$}
	\psfrag{J}[c][c]{\small$j$}
	\psfrag{Z}[c][c]{\small$Z/2$}
	\psfrag{R}[l][l]{\small Region A}
	\psfrag{T}[r][r]{\small Region B}
	\includegraphics[width=0.8\linewidth]{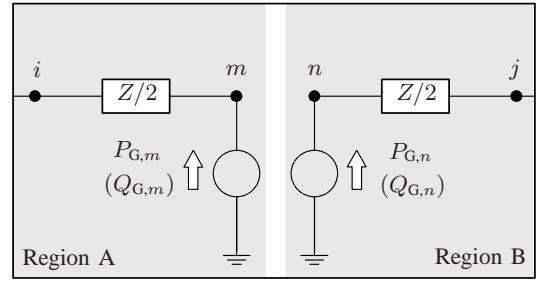}
	\caption[Tie line]{Decoupling model of a tie line (AC or DC) between nodes $i$ and $j$ with two auxiliary nodes $(m,n)$ and two auxiliary generators.}
	\label{fig:tieline}
\end{figure}

We cut the original line into two halves and add auxiliary nodes $(m,n)$ and auxiliary generators $(a,b)$ at both open ends, see Fig.~\ref{fig:tieline}. 
To guarantee a feasible power flow, the voltage must be equal at nodes $m$ and $n$. Furthermore, the generators must produce the same amount of power of the opposite sign. In the case of an AC tie line, we have complex voltage and both active and reactive power for boundary conditions:
\begin{subequations}\begin{align}
	|V_{\text{AC,}m}|&=|V_{\text{AC,}n}| \\ 
	\angle V_{\text{AC,}m}&=\angle V_{\text{AC,}n} \\ 
	P_{\text{G,}m}&=-P_{\text{G,}n} \\
	Q_{\text{G,}m}&=-Q_{\text{G,}n}.
	\end{align}\label{eq:consensus1}\end{subequations}
In the case of a DC tie line, we only have real voltage and active power to meet the constraints:
\begin{subequations}\begin{align}
	V_{\text{DC,}m}&=V_{\text{DC,}n} \\ 
	P_{\text{G,}m}&=-P_{\text{G,}n}.
	\end{align}\label{eq:consensus2}\end{subequations}
In a more compact form, the boundary conditions between region \textit{A} and \textit{B}, i.e. (\ref{eq:consensus1})-(\ref{eq:consensus2}), can be written as
\begin{align}
A_\text{A} x_\text{A} + A_\text{B} x_\text{B} =0.
\label{eq:consensus}\end{align}
\subsection{Problem Formulation}
We write
\begin{subequations}
	\begin{align}
	\underset{x_k,k\in\mathcal{R}}{\text{minimize}} & \quad \sum_{k\in \mathcal R^\text{}} f_{k}(x_{k}) \\
	\text{subject to} & \quad h_k(x_k) \leq 0, \quad \forall k\in\mathcal R, \\
	& \quad \sum_{k\in \mathcal R^\text{}} A_k x_k =0 ,\label{eq:consensus_2}
	\end{align}\label{eq:distOPF}
\end{subequations}
which is an equivalent OPF formulation as~(\ref{eq:OPF_2}), but in separable form. That is, each region $k$ has an objective $f_k$ depending only on local variables $x_k$. The same accounts for local constraints $h_k(x_k)\leq 0$. The equivalence to problem~(\ref{eq:OPF_2}) is guaranteed by consensus constraint~(\ref{eq:consensus_2}).

\subsection{Modified ADMM}\label{ch:Algorithms}
The general idea of ADMM, see also~\cite{Boyd2011}, is the following. Augmented regional OPFs are solved and the deviation of boundary variables from a fixed auxiliary variable $z$, which is information stemming from neighboring regions, is penalized. The regions then exchange information and $z$ is updated. The update is re-distributed to the local agents (areas) for a new OPF calculation until consensus between regions is achieved.
To this end, it is constructed an augmented Lagrangian of the form
\begin{align}
\mathcal{L}(x,z,\lambda) &= \sum_{k\in \mathcal R} \biggl\{ f_k(x_k)+\lambda^\top_k {A}_k (x_k-z_k) \notag\\
&+ \frac{\rho}{2}||{A}_k (x_k-z_k)||^2_{W} \biggr\} ,\label{eq:ADMM_Lag}
\end{align}
with penalty parameter $\rho\in\mathbb{R}$. With $n$ the number of consensus constraints, dual variables of such are denoted with $\lambda_k \in \mathbb{R}^{n\times 1}$, and $W \in\mathbb{R}^{n\times n}$ is a positive definite, diagonal weighting matrix, where each entry is related to one coupling constraint and can be dimensioned accordingly. 
In ADMM literature, $W$ is the identity matrix.
In three sequential steps, each of $x$, $z$, and $\lambda$ are optimized while the remaining two variables are fixed, we refer to~\cite{MeHue2019,Huebner2019,Huebner2019_Diss} for more details. An overview of the implemented ADMM is given in Algorithm~\ref{alg:ADMM}. Equation~(\ref{eq:ADMM_NLP}) solves the local nonlinear OPF's of each region to get optimized local variables $x_k$. In~(\ref{eq:ADMM_z1})-(\ref{eq:ADMM_z2}), information is sent to and gathered from neighboring regions to locally update auxiliary variables $z_k$. Then, dual variables $\lambda_k$ are updated depending on the current residuals.
An update rule on $\rho$ can be useful to enforce consensus. A $\rho_k\in\mathbb{R}$ is assigned to each region, which is increased depending on the local residual, see (\ref{eq:ADMM_rho}). If the residual has not decreased sufficiently compared to the previous iteration (indicator $0<\Theta\in\mathbb{R}<1$), the penalty is increased by a constant factor of $\tau\in\mathbb{R}>1$. 
The penalty parameter must be chosen carefully since it is widely known to be crucial for good convergence behavior~\cite{Mhanna2017}.

\begin{algorithm}
	\caption{ADMM}\label{alg:ADMM}
	\begin{algorithmic}[1]
		\State \textbf{Initialization:} Weighting matrix $W$, tolerance $\epsilon$; for all $k\in\mathcal{R}$: initial guesses $z_k$, penalty parameters $\rho_k=\rho$, dual variables $\lambda_k=\mathbf{0}$, local solutions $x_k=\boldsymbol{\infty}$, local residues $\Gamma_k=\infty$.
		\While{$||\sum_{k\in \mathcal R^\text{}} A_k x_k ||_\infty > \epsilon$}
		\State Solve for all $k\in\mathcal R$ the decoupled NLPs 
		\begin{subequations}\begin{align}
			\underset{x_k}{\text{min}} & \enskip f_k(x_k)+\lambda^\top_k {A}_k x_k + \frac{\rho_k}{2}||{A}_k (x_k-z_k)||^2_{W} \\
			\text{s.t.} & \enskip h_k(x_k) \leq 0 
			\end{align}\label{eq:ADMM_NLP}\end{subequations}
		\State For all $k\in\mathcal R$: Broadcast information on voltage-related constraints ($[\cdot]^\text{V}$) and power-related constraints ($[\cdot]^\text{S}$) to neighboring regions
		\begin{subequations}
			\begin{align}
		m^\text{V}_k = A^\text{V}_k x^\text{V}_k \\
		m^\text{S}_k = A^\text{S}_k x^\text{S}_k
		\end{align}\label{eq:ADMM_z1}
		\end{subequations}
		\State For all $k\in\mathcal R$: Update $z_k$ with information from each neighboring region $l$ for tie lines $kl$:
		\begin{subequations}
			\begin{align}
			z^\text{V}_{k,kl} &= 0.5(m^\text{V}_{k,kl} + m^\text{V}_{l,lk}) \\
			z^\text{S}_{k,kl} &= 0.5(m^\text{S}_{k,kl} - m^\text{S}_{l,lk})
			\end{align}\label{eq:ADMM_z2}
		\end{subequations}
		\State Update dual variables for all $k\in\mathcal R$  
		\begin{align}
		\lambda_k &\gets \lambda_k+\rho_k W {A}_k (x_k-z_k) \label{eq:ADMM_lam}
		\end{align}
		\State Calculate local residues and penalty parameter updates for all $k\in\mathcal R$ 
		\begin{subequations}\begin{align}
			\Gamma_k^+&=||A_k (x_k-z_k)||_\infty\\
			\rho_k &\gets \begin{cases}
			\mathcal \rho_k &\text{if}\enskip \Gamma_k^+\leq \Theta\Gamma_k \label{eq:ADMM_rho}\\
			\mathcal \tau\rho_k  &\text{otherwise}
			\end{cases}
			\end{align}\end{subequations}    
		
		\State Update $\Gamma_k\gets\Gamma_k^+$ for all $k\in\mathcal R$
		\EndWhile
\end{algorithmic}\end{algorithm}

\section{Case Study}\label{ch:Results}

\subsection{The European Network}\label{ch:case}

\begin{figure*}
	\centering
	\includegraphics[width=1\linewidth]{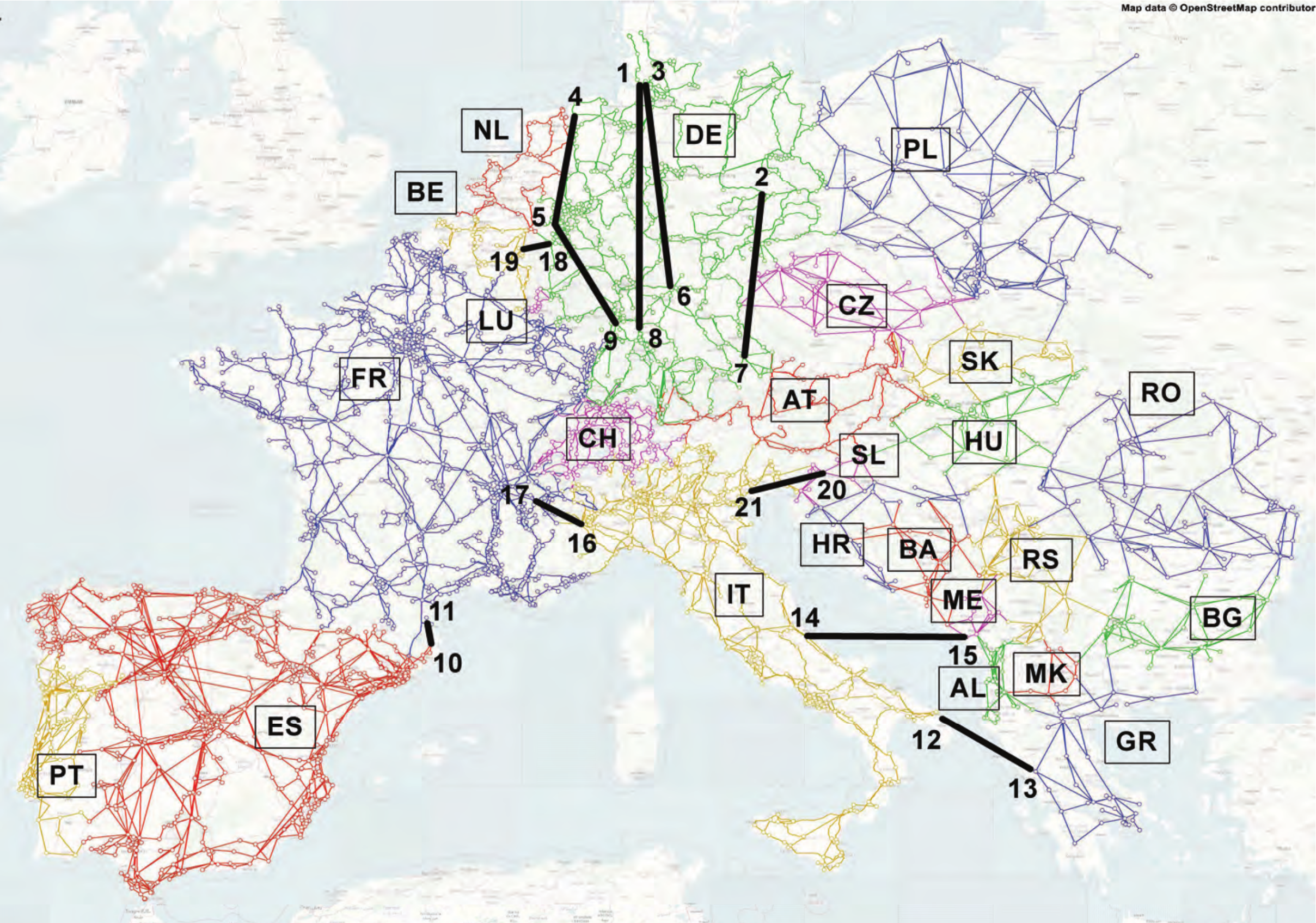}
	\caption{Model of the European transmission grid (220\:kV and 380\:kV). To apply a distributed OPF, the grid is decomposed into the colored areas, which correspond to the single countries indicated with a two-letter country code. DC lines are marked in black with the corresponding converter number.}
	\label{fig:netz2}
\end{figure*}

The European transmission grid is modeled based on the 2014 TYNDP public data set from ENTSO-E, which mainly covers the 220\:kV and 380\:kV voltage level. The model consists of 5707 buses, 8046 lines, 2134 generators, 3013 loads and 444 shunts. The given scenario includes a total load of 315.4\:GW and 75.3\:Gvar. Geographical locations are often hard to assign automatically due to unclear or anonymous station names. Nevertheless, a geographical model was achieved with intense manual effort, see Fig.~\ref{fig:netz2}. In central Europe, line corridors are additionally traced in detail. Eleven HVDC point-to-point connections are implemented -- indicated with black lines -- which results in 21 AC-DC converters. All converters are modeled with the above-mentioned standard VSC model for OPF studies, although older stations may be based on LCC technology. Inner-German converters 1-9 are rated with 1800\:MVA, all other converters with 900\:MVA. We use the same quadratic VSC loss function as in~\cite{MeHue2019}, leading to 1.1\:\% no-load losses and 1.85\:\% losses under full load.
The decoupled model consists of 24 regions with grid sizes between 8 nodes (Macedonia, MK) and 1535 nodes (France, FR). In total, the countries are connected by 210 AC and 6 DC cross-border tie lines. 

\subsection{Simulation Settings and Parameter Tuning}

We perform a reference solution (``centralized OPF'', \mbox{``OPF-c''}) by solving~(\ref{eq:OPF}) over the entire interconnected European grid. Since the problem is non-linear and non-convex, we do not expect a global optimum, but nevertheless a good local one. All OPF problems are solved with IPOPT~\cite{Waechter2006}.
We seek to minimize network losses and use the same costs of 50\:\euro~per MWh ($b_\text{G} = 50 \bigl[\frac{\text{\euro}}{\text{MWh}}\bigr]$) for each generator. Furthermore, we use a small quadratic cost function for all reactive power injections ($a_\text{Q} = 0.001 \bigl[\frac{1}{\text{Mvar}^2}\frac{\text{\euro}}{\text{h}}\bigr]$). The base power for p.u.-values is 100\:MVA.

\begin{figure}[tbp]
	\centering
	\includegraphics[width=0.98\linewidth]{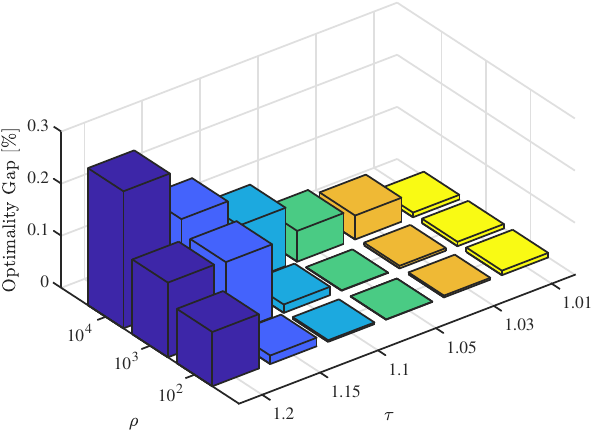}
	\caption{Optimality gap of each simulation in a parameter sweep of $\rho$ and $\tau$.}
	\label{fig:cost3d}
\end{figure}

\begin{figure}[tbp]
	\centering
	\includegraphics[width=0.98\linewidth]{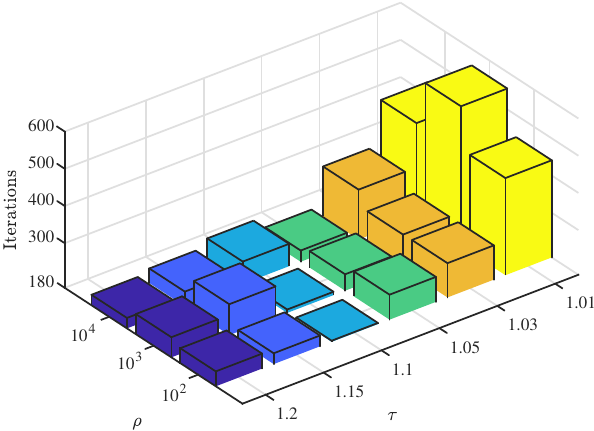}
	\caption{Iterations of each simulation in a parameter sweep of $\rho$ and $\tau$.}	
	\label{fig:iterations3d}
\end{figure}

We chose a convergence criteria of $\epsilon=0.001$, that is, 0.1\:\% voltage deviation and 10\:kVA power deviation. We set $\Theta=0.99$ and penalty parameters $\rho$ and $\tau$ are chosen after a parameter sweep. The two decisive criteria are optimality gap and number of iterations, which are shown in Fig.~\ref{fig:cost3d} and Fig.~\ref{fig:iterations3d}, respectively. 
In general, large penalties prevent near-optimal solutions, since the objective function is dominated by the penalty instead of the original cost function. As a consequence, it can be observed that the optimality gap is increased with both $\rho$ and $\tau$, that is, if the penalty is already large at the beginning or ramped up too fast. On the other hand, the convergence criterion is cross-border feasibility among the regions. Thus, if consensus is strongly enforced with high penalties, convergence is achieved faster. In Fig.~\ref{fig:iterations3d}, we observe a dependency between convergence speed and penalty ramping $\tau$. Convergence is achieved for all cases under 250 iterations if $\tau>1.03$. Fastest convergence is achieved with $\tau=1.1$ and $\rho\in\{10^2,10^3\}$. Thus, we choose $(\tau,\rho)=(1.1, 10^2)$ for the remainder since the optimality gap is acceptable (0.016\:\%). We emphasize that the sweep of two parameters is straightforward. Furthermore, there is a certain range where the optimization leads to acceptable speed and optimality. 
For the above analysis, we chose matrix $W$ from \cite{MeHue2019} to show the generality of the chosen weights. That is, voltage-related entries are set to 100, and power-related entries are set to 1. However, convergence can further be improved with empirical studies on certain variable weights. For example, we reduce the number of iterations from 189 to 166, if we increase entries in $W$ related to DC-power from 1 to 10. We use this setting in the following case study.


\subsection{Case Study Results}

\begin{figure}[tbp]
	\centering
	\includegraphics[width=0.98\linewidth]{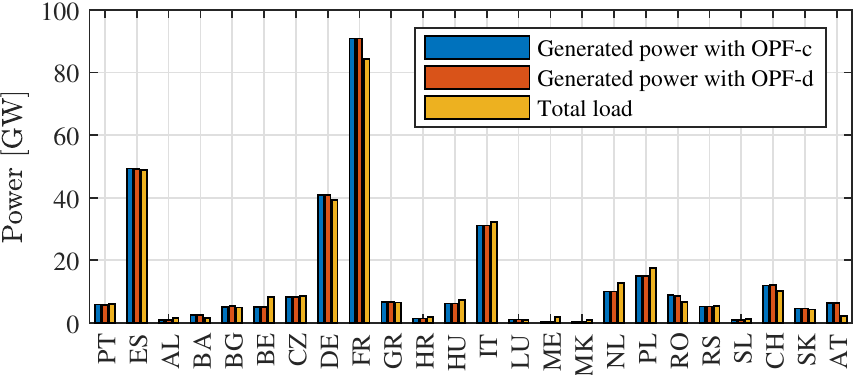}
	\caption{Comparison of total generated power and total load for each country with centralized (``OPF-c'') and distributed (``OPF-d'') calculation.}
	\label{fig:comp_ENTSO1}
\end{figure}

\begin{figure}[tbp]
	\centering
	\includegraphics[width=0.97\linewidth]{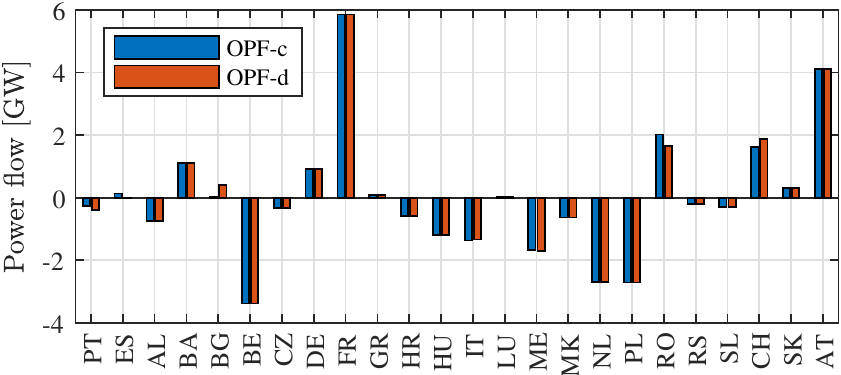}
	\caption{Cumulated power flow over tie lines for each country. Positive value: net power export, negative value: net power import.}
	\label{fig:comp_ENTSO2}
\end{figure}

In Fig.~\ref{fig:comp_ENTSO1}, we compare the power balance between centralized and distributed calculation for each country, where total generated power and total load are depicted. The difference between centralized and distributed OPF is marginal. As a result, the cumulated power flow over tie lines is very similar as well, as shown in Fig.~\ref{fig:comp_ENTSO2}. A positive value indicates an ``export country'', a negative value an ``import country''.

\begin{figure}[tbp]
	\centering
	\includegraphics[width=0.99\linewidth]{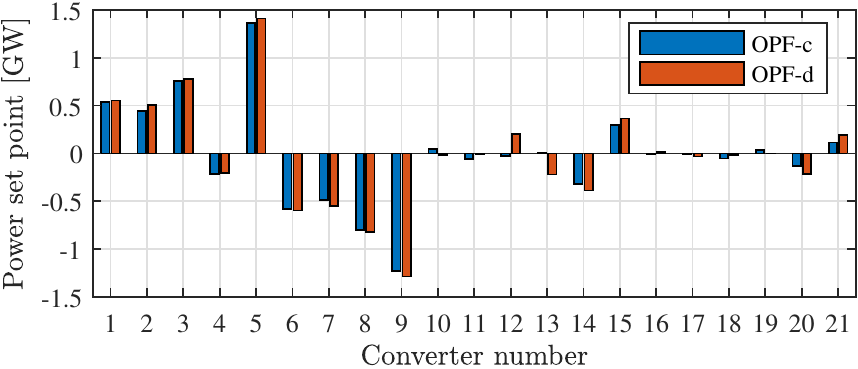}
	\caption{Active power set points of AC-DC converters with centralized and distributed calculation. Positive value: power transfer from AC to DC.}
	\label{fig:PsConv}
\end{figure}

In Fig.~\ref{fig:PsConv}, the active power set point is shown for each converter. It is notable, that the distributed solution is in general close to the centralized one. However, we observe small deviation for instance at converters 12 and 13, which is a tie line between Italy and Greece. Here, a power transfer of about 200\:MW towards Greece is calculated with the distributed method, while the tie line is almost inactive with a centralized calculation.

\begin{figure}[tbp]
	\centering
	\includegraphics[width=0.99\linewidth]{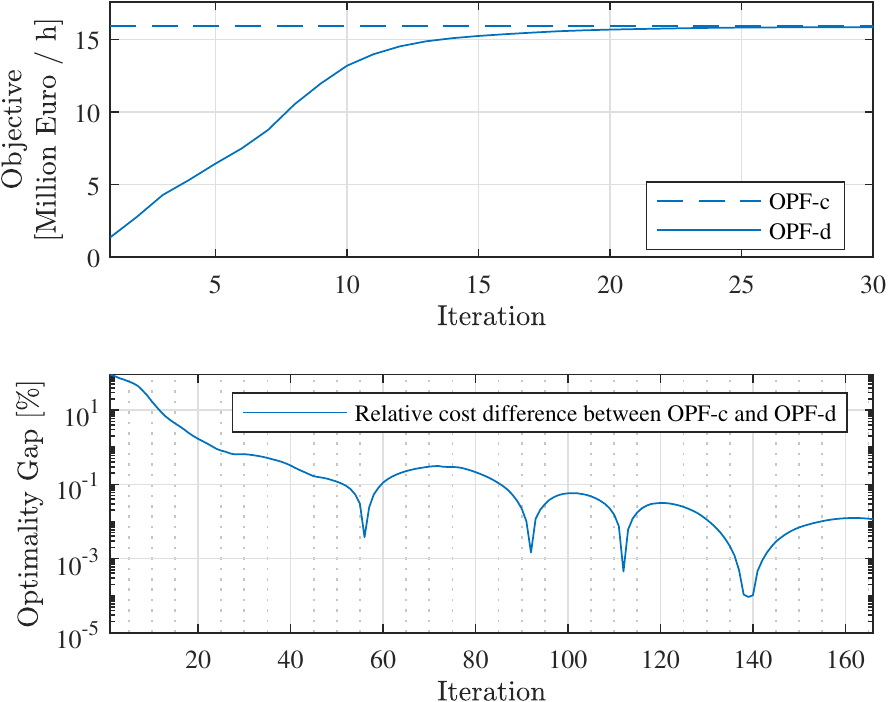}
	\caption{Cost difference (sub-optimality) between centralized and distributed OPF. Top: absolute objective value. Bottom: relative difference.}
	\vspace{3pt}
	\label{fig:objective}
\end{figure}

However, the decisive measure of whether a set point is ``worse'' than another, is the effect on objective values. In Fig.~\ref{fig:objective}, we show the sub-optimality of a distributed solution. The top figure shows the absolute objective value, that is, the total operating cost of the system. During the first 30 iterations, the distributed solution approaches the centralized one of 15.2 Million Euros per hour very closely. In the bottom figure, the cost difference is depicted in detail. The final deviation after convergence remains around 0.02\:\% or 3000 Euro, which is remarkably low. Thus, the above-mentioned deviation in converter power set points has a negligible influence on operating costs.

\begin{figure}[tbp]
	\centering
	\includegraphics[width=0.99\linewidth]{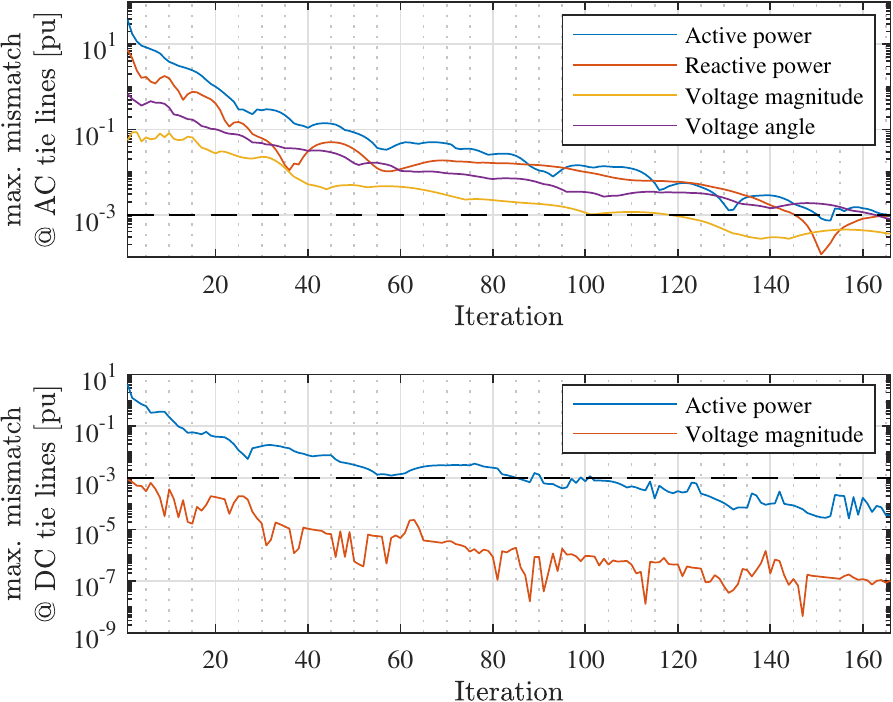}
	\caption{Maximal mismatch of coupling variables between neighboring countries. Top: AC tie lines. Bottom: DC tie lines.}
	\label{fig:residual}
\end{figure}

%

The second major criterion in distributed optimization is the feasibility. Separately from each other, two neighboring regions calculate voltage and transferred power at a coupling bus. As a result, the two obtained solutions should be close to identical in order to allow for a physically feasible load flow. In Fig.~\ref{fig:residual}, we show the maximal mismatch between two neighbors over the iterations. In the case of an AC tie line (top figure), we have four coupling variables. In the beginning, active and reactive power mismatches can be quite large compared to the relative voltage mismatches, that is, over 1\:GVA or $10^1$\:pu. Toward the end of optimization, voltage angle and active power tend to be more difficult to converge than the other two variables. This is supposedly since reactive power and voltage magnitude can be adapted locally more easily. In the case of a DC tie line, only voltage magnitude and active power must be coupled. Here, the maximal mismatches satisfy the convergence criterion after 100 iterations.

\begin{figure}[tbp]
	\centering
	\includegraphics[width=0.98\linewidth]{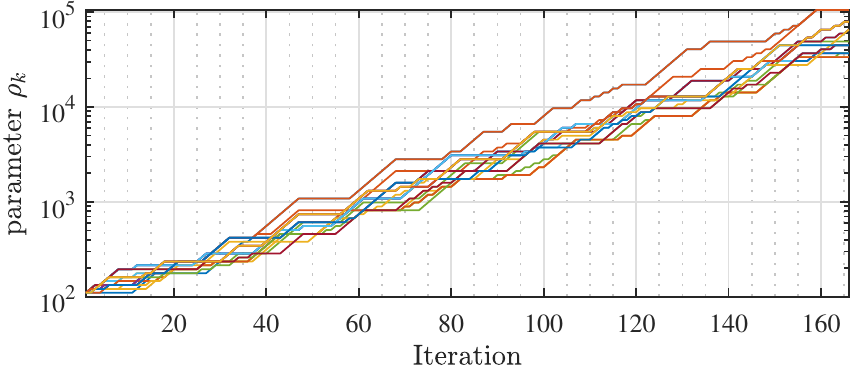}
	\caption{Penalty parameter $\rho_k$ for each region $k$.}
	\label{fig:rho}
\end{figure}

Lastly, we show the penalty parameter $\rho_k$ for each region $k$ in Fig.~\ref{fig:rho}. It is increased from 100 to a maximum of 10000 for certain regions. Interestingly, the highest penalties are imposed on Spain and Portugal.

\section{Discussion \& Outlook}\label{ch:Conclusion}

This work shows that distributed optimization is applicable to large-scale realistic transmission grids. We present the first case study in a grid of such geographical dimensions and total load. With over 200 tie lines and many small regions, the partitioning is mathematically challenging but can be implemented in the existing regulatory and political framework, since it corresponds to the country borders. At the same time, data exchange is very low and requirements on each system operator's privacy can be fulfilled. Nevertheless, fast convergence is achieved with a cost difference of 0.02\:\% compared to a centralized optimization. We use both AC and DC cross-border tie lines and a multi-terminal DC system, while the model is designed for meshed DC systems as well. In this work, a conventional AC-DC OPF is solved to minimize network losses. Thermal branch flow limits should be incorporated in the next step. Furthermore, generator redispatch and N-1 security will play an important role in the future, which has been optimized distributively in a small grid in~\cite{Huebner2019}, but is yet to be demonstrated in large-scale.


\bibliographystyle{IEEEtran}
\bibliography{DistributedENTSOE}

\end{document}